\documentclass[preprint2]{aastex}

\shorttitle{HST/FOS Spectral Mapping of V2051 Ophiuchi in a Low State}
\shortauthors{Saito \& Baptista}

\begin{document}

\title{HST/FOS Spectral Mapping of V2051 Ophiuchi in a Low State}

\author{R. K. Saito and R. Baptista} \affil{Departamento de
F\'{\i}sica, Universidade Federal de Santa Catarina, Brazil}

\altaffiltext{}{Departamento de F\'{\i}sica, Universidade Federal de
Santa Catarina, Trindade, 88040-900, Florian\'opolis, SC, Brazil}

\begin{abstract}
We report a study of the spectra and structure of the accretion disk
of the dwarf nova V2051 Ophiuchi while the star was in an unusual
faint brightness state during 1996. The differences between the UV
resonant lines and continuum disk surface brightness distributions
indicate a vertically extended disk with the emission from these lines
arising from the upper atmospheric layers. Distinct emission along the
stream trajectory suggests the occurrence of gas stream
overflow. Spatially resolved spectra show that the lines are in
emission at all disk radii. The Balmer decrement becomes shallower
with increasing radius. The FWHMs of the emission lines show
differences with respect to that expected for a gas in Keplerian
rotation and the line intensities drop with a radial dependency of
$I\propto R^{-1.78}$.  The uneclipsed light contributes about 5 - 10
per cent of the total flux, and its spectrum is dominated by a Balmer
jump and strong lines in emission. Broad absorption bands, possibly
due to Fe\,II, are present in the spectra of the disk side farther
away from the secondary star, suggesting it arises from absorption by
a extended gas region above the disk; the differences between the
spectra of the hemisphere farther from and nearer to the secondary
star are interpreted in terms of chromospheric emission from a disc
with a non-negligible opening angle (limb brightening effect). Stellar
atmosphere model fits to the extracted white dwarf spectrum lead to a
temperature $T_{WD}= 9500\,^{+2900}_{-1900}\,K$ and a distance of
$d = 67\,^{+22}_{-25}\,pc$ if the inner disk is opaque, or $d =
92\,^{+30}_{-35}\,pc$ if the inner disk is optically thin.
\end{abstract}

\keywords{stars: cataclysmic variables -- accretion disks, dwarf
               novae -- individual: V2051~Oph}

\section{Introduction}

V2051 Ophiuchi was discovered by Sanduleak (1972). It is a dwarf nova,
a compact binary in which mass is fed to a white dwarf (the primary
star) by a Roche lobe filling late-type star (the secondary) via an
accretion disk. V2051 Oph shows recurrent but sparse outbursts in
which the disk expands and brightens by 2-3 magnitudes during 1-3 days
($B\simeq 13 \;mag$ at maximum, Bateson 1980; Warner \& Cropper 1983;
Warner \& O'Donoghue 1987) as a consequence of a sudden increase in
mass accretion through the disk. In the quiescent, low-mass accretion
state, the white dwarf and the bright spot (formed by the impact of
the infalling gas stream with the outer edge of the disk) dominate the
light from the binary at optical and ultraviolet wavelengths, and the
optical spectrum shows double-peaked H\,I and He\,II emission lines
which exhibit the classical rotational disturbance effect during
eclipse (Cook \& Brunt 1983; Watts et al. 1986).

The binary is seen at a high inclination angle ($i=83\degr$), which
leads to deep eclipses ($\Delta B \simeq$ 2.5 mag) in the light curve
every 1.5 hours when the white dwarf, accretion disk and bright spot
are occulted by the mass-donor star. This allows the emission from the
different light sources to be distinguished and spatially resolved
studies to be performed, making V2051 Oph an excellent laboratory for
the study of accretion physics. In particular, it is well suited for
the application of indirect imaging techniques to resolve the disk
emission both in position (eclipse mapping, Horne 1985) and velocity
(Doppler tomography, Marsh \& Horne 1988).

V2051 Oph displays large amplitude flickering activity (up to 30 per
cent of the total light in the optical), which is responsible for a
variety of eclipse morphologies (e.g., Warner \& Cropper 1983) and
usually hampers the measurement of white dwarf and bright spot eclipse
phases.  Baptista et al. (1998) caught the star in an unusual faint
brightness state ($B\simeq 16.2 \;mag$) in which the mass accretion
rate and flickering activity were significantly reduced, allowing a
clean view of the white dwarf at disk center. They derived a mass
ratio of $q=0.19 \pm 0.03$ and an inclination of $i=83.\degr3 \pm
1.\degr4$.

Vrielmann et al. (2002) applied eclipse mapping techniques to
multicolor data of V2051 Oph to find that the disk is brighter in the
front side (the hemisphere nearer to the secondary) and interpreted
this behavior as caused by enhanced emission from the bright
spot. They also estimated a distance of $d=146 \pm 20\,pc$ for the
system.

Kiyota \& Kato (1998) observed V2051 Oph in superoutburst and detected
superhumps in the light curve of the object, confirming its
classification as an SU UMa star -- a sub-class of dwarf novae with
occasional superoutbursts $\sim 0.7$ mag brighter and $\sim 5$ times
longer than an ordinary outburst. However, due to the complex behavior
observed in V2051 Oph, Warner (1996) suggested a possible alternative
classification of it as a polaroid, an intermediate polar with a
synchronized primary.

\begin{figure*}
\includegraphics[bb=3cm 1cm 20cm 16cm,angle=-90,scale=0.65]{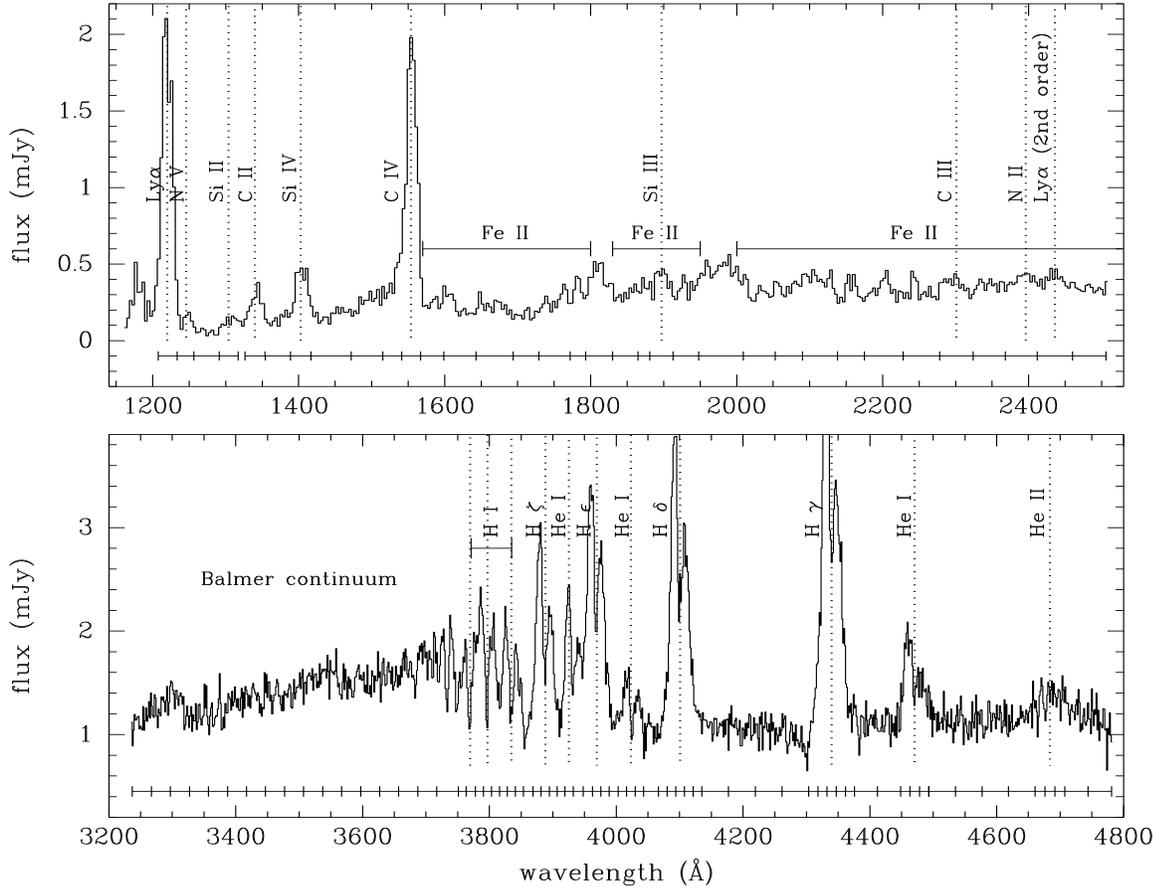}
 \caption{Average out-of-eclipse UV (G160L; top ) and optical (G400H;
bottom) HST spectra of V2051~Oph.  The phase ranges are +0.10 to +0.20
and +0.075 to +0.080 cycles, respectively, for the UV and optical
spectra. The observations correspond to the eclipse cycles 109,988
(optical) and 109,989 (UV) according to the linear ephemeris of
Baptista et al. (2003). Emission (vertical dotted lines) and
absorption (horizontal lines) features are labeled. Horizontal tick
marks indicate the passbands used to extract light curves (34 in the
UV and 68 in the optical).}
 \label{spec_med}
\end{figure*}

Here we report the results of an optical and UV spectral mapping
experiment of V2051 Oph in its faint brightness state of 1996
(e.g. Baptista et al. 1998). The details of the data analysis are
given in Section 2. In Section 3 we investigate the disk spatial
structure in the lines and continuum, and we present spatially
resolved spectra of the disk, gas stream and uneclipsed light. The
results are discussed in Section 4 and summarized in Section 5.

\section{Data analysis}

\subsection{Observations}

The Faint Object Spectrograph (FOS) onboard the Hubble Space Telescope
(HST) was used to secure high-speed spectroscopy covering two
consecutive eclipses of V2051 Oph on 1996 January 29 while the star
was in an unusual low brightness state ($B\simeq 16.2$ mag). The
observations are summarized in Table 1. The first run covers a short
phase range around the egress of an eclipse in the optical (G400H
grating), and the second run covers the next eclipse in the UV (G160L
grating). The reader is referred to Baptista et al. (1998) for a
detailed description of the observations and data reduction
procedures.

\begin{table*}
\begin{center}
\caption{Journal of the Observations.\label{tbl-2}}
\vspace{12pt}
\begin{tabular}{c c c c c c}
\tableline\tableline
Run & Start & Eclipse & Number of & Spectral Range & Phase Range  \\
    & (UT)  & Cycle   & Exposures & \AA            & (cycles)     \\  
\tableline
  H1  & 19:03 & 109,988 & 113       & $3226-4781$ & $+0.01, +0.08$ \\
  H2  & 20:24 & 109,989 & 693       & $1150-2507$ & $-0.09, +0.34$ \\
\tableline
\end{tabular}
\end{center}
\end{table*}

The average out-of-eclipse UV spectrum is shown in
Fig.\,\ref{spec_med} (top). It displays emission lines of Ly$\alpha$
$\lambda 1216$ (mostly geo coronal), N\,V $\lambda 1240,1243$, Si\,II
$\lambda 1300$, C\,II $\lambda 1336$, Si\,IV $\lambda 1394,1403$,
C\,IV $\lambda 1549,1551$, Si\,III $\lambda 1892$, C\,III $\lambda
2297$, as well as broad absorption bands reminiscent of those seen in
OY Car (Horne et al. 1994), which were interpreted as a blend of
Fe\,II lines. There is no evidence of He\,II $\lambda 1640$
emission. The position of the second-order $\rm Ly\alpha$ emission is
illustrated in Figure \ref{spec_med} (top). The bottom panel of Figure
\ref{spec_med} shows the average out-of-eclipse optical spectrum.  It
displays the Balmer continuum in emission, as well as He\,I, He\,II
and strong Balmer emission lines. The optical emission lines show a
clear double-peaked profile. The blue peak of the lines is stronger
than the red peak because part of the disk side contributing
redshifted line emission is still occulted by the secondary star at
the phase range used to compute the average spectrum.

\subsection{Light-curve construction}

The UV spectra were divided into a set of 34 narrow passbands, with 22
continuum passbands 19-60\,\AA\ wide and 12 passbands for 10 lines
(Fig.\,\ref{spec_med}). The blue end of the spectra (shortward of the
$Ly\alpha$ line) was not included in this analysis because of the very
low count rate [and, consequently, low signal-to-noise ratio (S/N)] of
the data in this spectral region. The emission lines were sampled in a
single velocity bin 5000 or 6000$\,km\,s^{-1}$ wide (centered at the
rest wavelength, $v=0\,km\,s^{-1}$), except for the C\,IV line, which
was separated in a bin centered at the rest wavelength and
velocity-resolved bins for the blue and red wings of the line (all
bins with $\Delta v= 5000\,km\,s^{-1}$)\footnote{The choice of the
number (or width) of the passbands was constrained by the
signal-to-noise (S/N) ratio of the resulting light curves. Narrower
passbands lead to light curves too noisy to produce reliable eclipse
maps.}. The optical spectra were divided into a set of 68 passbands,
with 32 continuum passbands 15-42\,\AA\ wide and 36 passbands for 11
lines. The emission lines were resolved in velocity bins
$1000\,km\,s^{-1}$ wide, with one bin centered at the rest
wavelength. The systemic velocity of V2051~Oph ($\gamma\simeq
40\,km\,s^{-1}$; Watts et al. 1986) is both rather uncertain and much
smaller than the width of the passbands and was neglected. For those
passbands including emission lines the light curve comprises the total
flux at the corresponding bin, with no subtraction of a possible
continuum contribution.

Light curves were extracted for each passband by computing the average
flux on the corresponding wavelength range and phase-folding the
results according to the linear ephemeris of Baptista et al. (2003),

\begin{equation}
\rm T_{mid}(BJDD)=2\,443\,245.97752+0.062\,427\,8634~E,
\end{equation} 
where $\rm T_{mid}$ gives the inferior conjunction of the white dwarf.

Light curves at two selected continuum passbands and for the C\,IV and
H$\gamma$ lines can be seen in Figs.\,\ref{mapas_uv} and
\ref{mapas_op}, respectively. The incomplete eclipse phase coverage of
the H1 run is clear (Fig.\,\ref{mapas_op}).

\subsection{Eclipse mapping}

Maximum-entropy eclipse mapping techniques (Horne 1985, Baptista \&
Steiner 1993) were used to solve for a map of the disk brightness
distribution and for the flux of an additional uneclipsed component in
each band. The reader is referred to Baptista (2001) for a recent
review on the eclipse mapping method.

As our eclipse map we adopted a flat grid of $51\times51$ pixels
centered on the primary star with side $2\,R_{L1}$, where $R_{L1}$ is
the distance from the disk center to the inner Lagrangian point. The
eclipse geometry is defined by the inclination $i$ and the mass ratio
$q$.  The mass ratio $q$ defines the shape and the relative size of
the Roche lobes. The inclination $i$ determines the shape and
extension of the shadow of the secondary star as projected onto the
orbital plane. In this paper we adopted the values derived by Baptista
et al. (1998) , $\rm q=0.19$ and $\rm i=83\degr$, which correspond to
an eclipse phase width of $\Delta\phi= 0.0662$\,cycle. This
combination of parameters ensures that the white dwarf is at the
center of the map.

For the reconstructions we adopted the default of limited azimuthal
smearing of Rutten et al. (1992), which is better suited for
recovering asymmetric structures than the original default of full
azimuthal smearing (e.g. Baptista et al. 1996). We used a radial blur
width $\Delta r = 0.0157 R_{L1}$ and an azimuthal blur width $\Delta
\theta = 30 \degr$.

The statistical uncertainties in the eclipse maps were estimated with
a Monte Carlo procedure (e.g., Rutten et al. 1992).  For a given
narrow band light curve, a set of 20 artificial light curves was
generated in which the data points were independently and randomly
varied according to a Gaussian distribution with standard deviation
equal to the uncertainty at that point. The light curves were fitted
with the eclipse-mapping code to produce a set of randomized eclipse
maps. These were combined to produce an average map and a map of the
residuals with respect to the average, which yields the statistical
uncertainty at each pixel. The uncertainties obtained with this
procedure were used to estimate the errors in the derived radial
brightness distributions, as well as in the spatially resolved
spectra.

The Appendix presents the results of eclipse mapping simulations
addressing the reliability of disk surface brightness reconstructions
from light curves of incomplete phase coverage, such as those of run
H1.

Light curves and respective eclipse maps at selected passbands for the
continuum and for the strongest emission lines are shown in
Figs.\,\ref{mapas_uv} and \ref{mapas_op}, respectively. Contour curves
in these Figures enclose the regions of the eclipse maps at the
3$\,\sigma$ and 5$\,\sigma$ levels of statistical significance.


\section{Results}

\subsection{Disk structure}

Maps of the disk surface brightness distributions, calculated by the
maximum entropy eclipse mapping method, allow us to obtain spatially
resolved information about the disk emission. In this section we
discuss the structures in the eclipse maps of the strongest lines in
the spectrum, as well as in selected continuum passbands.

 \begin{figure*}
 \centering
 \includegraphics[bb=5cm 2cm 20cm 14cm,angle=-90,scale=.65]{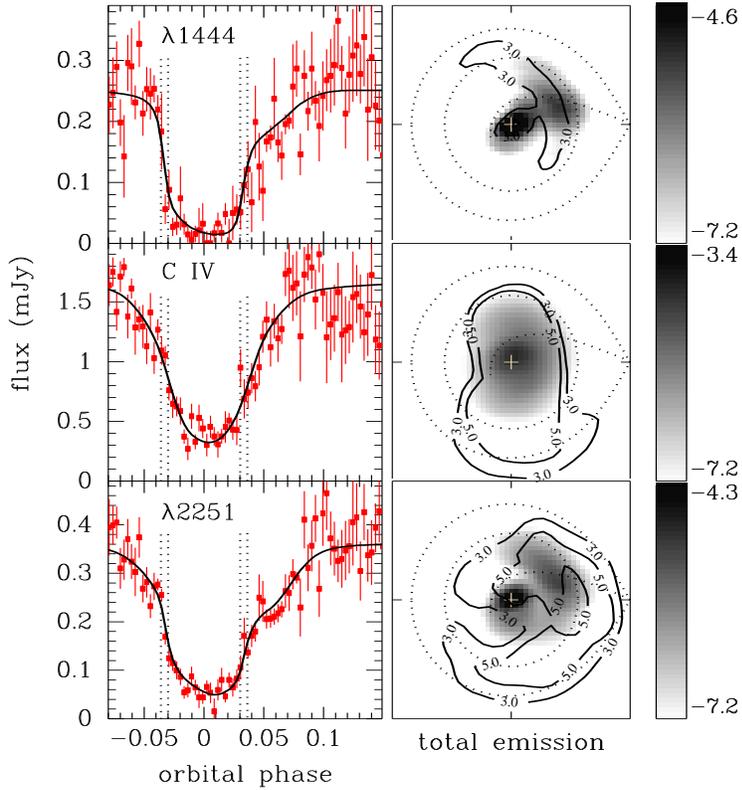}
     \caption{Data and model light curves and eclipse maps for the
C\,IV line center ($\Delta{v}=5000\,km\,s^{-1}$) and for two selected
continuum passbands in the UV. Left: Data (dots with error bars) and
model (solid lines) light curves. Dotted lines mark the contact phases
of the white dwarf eclipse. Right: Corresponding eclipse maps in a
logarithmic gray scale.  Brighter regions are shown in black, fainter
regions in white. Contour curves enclose the eclipse map regions at
the 3$\,\sigma$ and 5$\,\sigma$ levels of statistical significance. A
cross marks the center of the disk; dotted lines show the Roche lobe,
the gas stream trajectory, and a disk of radius $0.56\;R_{L1}$; the
secondary is to the right of each map and the stars rotate
counter-clockwise. The vertical bars indicate the logarithmic
intensity level of the gray scale in each map.}
        \label{mapas_uv}
  \end{figure*}

Fig.\,\ref{mapas_uv} shows the light curves and respective eclipse
maps for the C\,IV line center and for two selected UV continuum
passbands. The S/N ratio of the light curves decreases towards the
blue end of the spectrum. Accordingly, the statistical significance of
the eclipse maps is lower in this spectral region. Thus, the brightest
parts of the $1444$\,\AA~continuum map are only significant at the
$3\,\sigma$ level, while the eclipse maps in the red side of the
spectrum are typically significant at the $5\,\sigma$ confidence
level.

The C\,IV line center light curve shows an inverse ``Gaussian''-shaped
eclipse with a minor asymmetry at egress and no sharp breaks in
slope. Accordingly, the eclipse map shows a broad, smooth, and fairly
symmetrical brightness distribution centered on the white dwarf, with
no clear sign of the bright spot.

In contrast to what is observed in the C\,IV line center, the
continuum light curves show sharp breaks in slope at the ingress and
egress phases of the white dwarf plus a conspicuous asymmetric egress
shoulder, indicating that there is significant additional emission
from the disk side containing the bright spot. The corresponding
eclipse maps display pronounced emission from the compact white dwarf
at disk center plus an asymmetric structure in the disk side that is
moving away from the secondary star (the upper hemisphere of the
eclipse maps in Fig.\,\ref{mapas_uv}) that can be associated with the
gas stream emission.

The presence of bright compact structures in the continuum maps and
their absence in the C\,IV line map, in addition to the observed
spectrum of the uneclipsed component (see Section 4.4), are
indications of the vertical extension and large optical depth of the
gas from which the C\,IV line originates. Our interpretation is that
the C\,IV line emission arises in a large gas region surrounding the
disk. This region needs to be vertically extended and optically thick
at this wavelength in order to veil the strong emission from the
underlying white dwarf and bright spot. This is in line with
expectations for an emission line produced by resonant scattering in a
disk chromosphere. A similar result was found for the nova-like
variable UX UMa (Baptista et al. 1995).

\begin{figure*}
  \centering
    \includegraphics[bb=5cm 9.5cm 20cm 14cm,angle=-90,scale=.65]{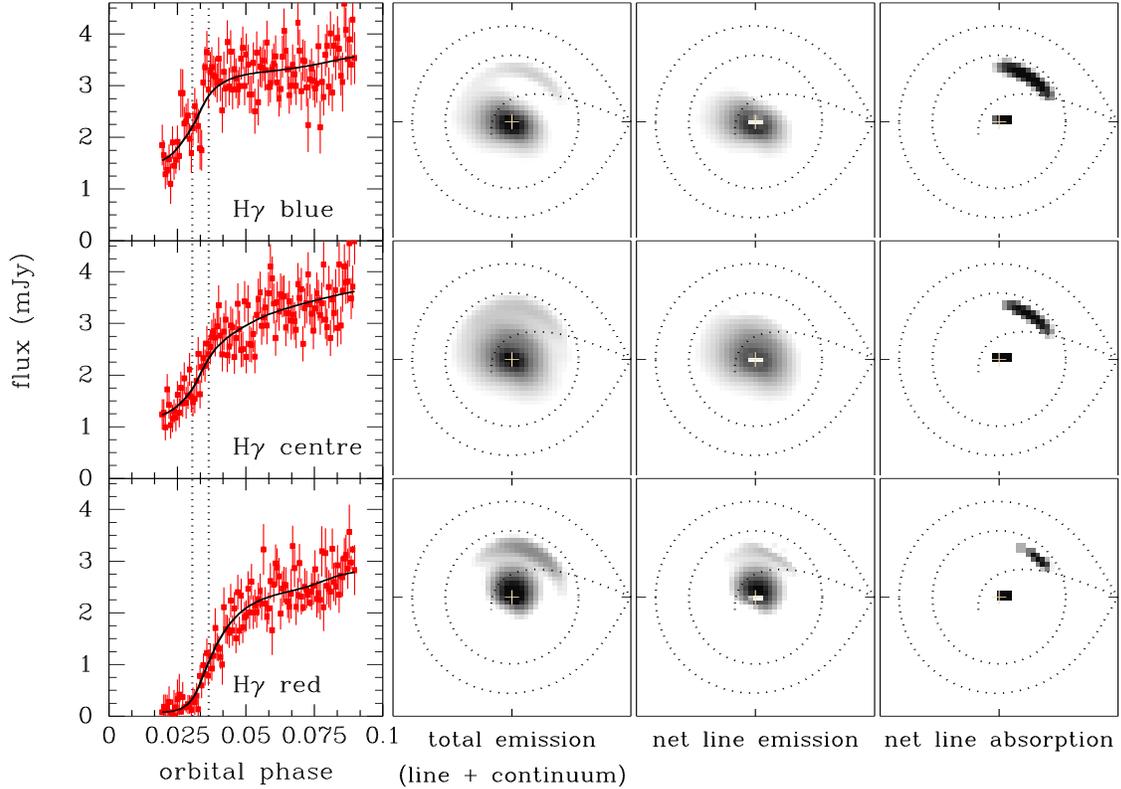}
       \caption{Left: Data (dots with error bars) and corresponding
  model (solid lines) light curves for the $H\gamma$ line center
  passband ($v=0\,km\,s^{-1}, \Delta{v}=1000\,km\,s^{-1}$) and for its
  blue and red line wings ($\Delta{v}=1000\,km\,s^{-1}$). Second from
  left: Corresponding eclipse maps (total emission). Third from left:
  Eclipse maps of the net line emission. Right: Eclipse maps of the
  net line absorption in each case. The gray scale is the same in each
  column. The notation is similar to that of Fig.\,\ref{mapas_uv}.}
          \label{mapas_op}
\end{figure*}

The eclipse map at $2251$\,\AA~is representative of most of the
continuum maps. Aside from the asymmetry caused by the bright spot and
gas stream emission, it shows that another brightness asymmetry,
namely, the disk hemisphere closest to the secondary star (the
``front'' disk side) is systematically brighter than the disk side
farther away from the secondary star (hereafter called the ``back''
disk side). A similar effect was observed by Vrielmann et al. (2002),
who interpreted it as a consequence of azimuthally-smeared bright spot
emission. Our results will lead to a different interpretation (see
Section 4.1).

Fig.\,\ref{mapas_op} shows light curves and eclipse maps for the
$H\gamma$ line center ($v=0\,km\,s^{-1}$) and the blue
($v=-1000\,km\,s^{-1}$) and red ($v=+1000\,km\,s^{-1}$) line wing
passbands. The velocity-resolved $H\gamma$ line light curves show the
expected behavior for the eclipse of gas rotating in the prograde
sense (rotational disturbance), with the blue wing of the line
reappearing from eclipse earlier than the red wing. The phase of
maximum derivative in the three light curves coincides with the white
dwarf egress phase, indicating that the maxima of the distributions
are at the disk center. The light curves show deep eclipses with
smooth and long lasting egresses.  As a consequence, the resulting
maps show fairly broad brightness distributions centered at the white
dwarf position with an asymmetric bright source in the quadrant
containing the bright spot. Simulations show that, in spite of the
incomplete phase coverage of the optical light curves, it is possible
to derive a fairly good reconstruction of the position and intensity
of the white dwarf and bright spot in the corresponding eclipse maps
(see the Appendix).

Fig.\,\ref{mapas_op} (third and fourth columns) shows the net line
emission and the net line absorption in each case. Net line
emission/absorption maps are obtained by combining continuum eclipse
maps on the short- and long-wavelength sides of the target emission
line and by subtracting the derived average continuum map from each of
the velocity-resolved line maps. Positive intensities in the resulting
maps (third column, black and gray) signal the regions where the line
appears in emission, while negative intensities (fourth column, black
and gray) trace the regions where the line is in absorption. The net
emission/absorption maps reveal that the Balmer lines are in emission
over most of the accretion disk, but appear in absorption at the disk
center and also in the region of the bright spot at the disk rim.

\subsection{Spatially resolved spectra}

  \begin{figure}
  \centering
  \includegraphics[bb=1cm 1cm 20cm 21cm,angle=-90,scale=.4]{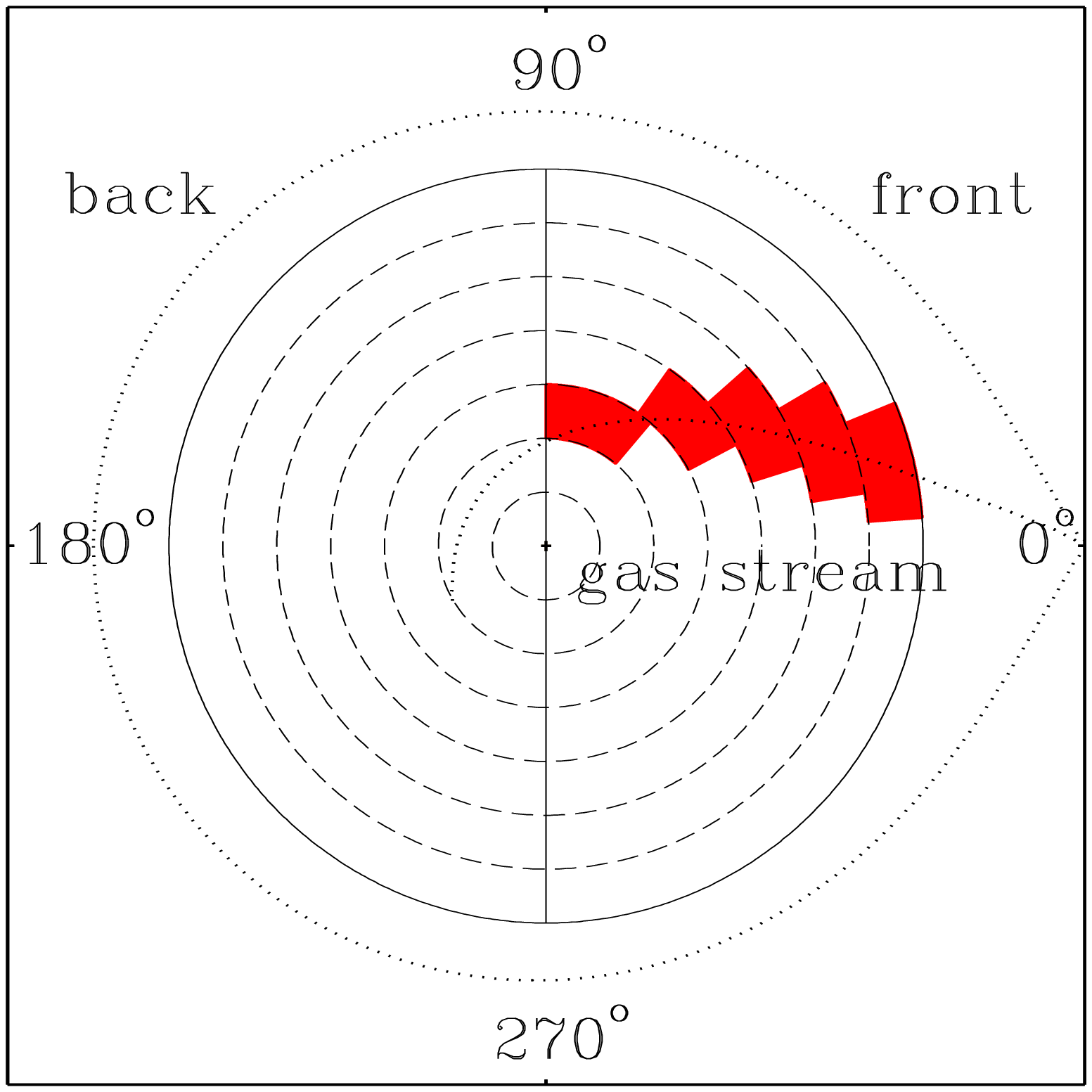}
     \caption{Schematic diagram showing the regions defined as
``front'', ``back'' and ``gas stream''. Dashed lines mark the annular
regions of width $0.1\,R_{L1}$ used to extract spatially resolved
spectra. Dotted lines show the projection of the primary Roche lobe
onto the orbital plane and the gas stream trajectory. Azimuths are
measured with respect to the line joining both stars and increase
counter-clockwise. Four reference azimuths are labeled.}
        \label{mapa_esq}
  \end{figure}

  \begin{figure*}
  \centering
  \includegraphics[bb=6cm 12cm 20cm 14cm,angle=-90,scale=.74]{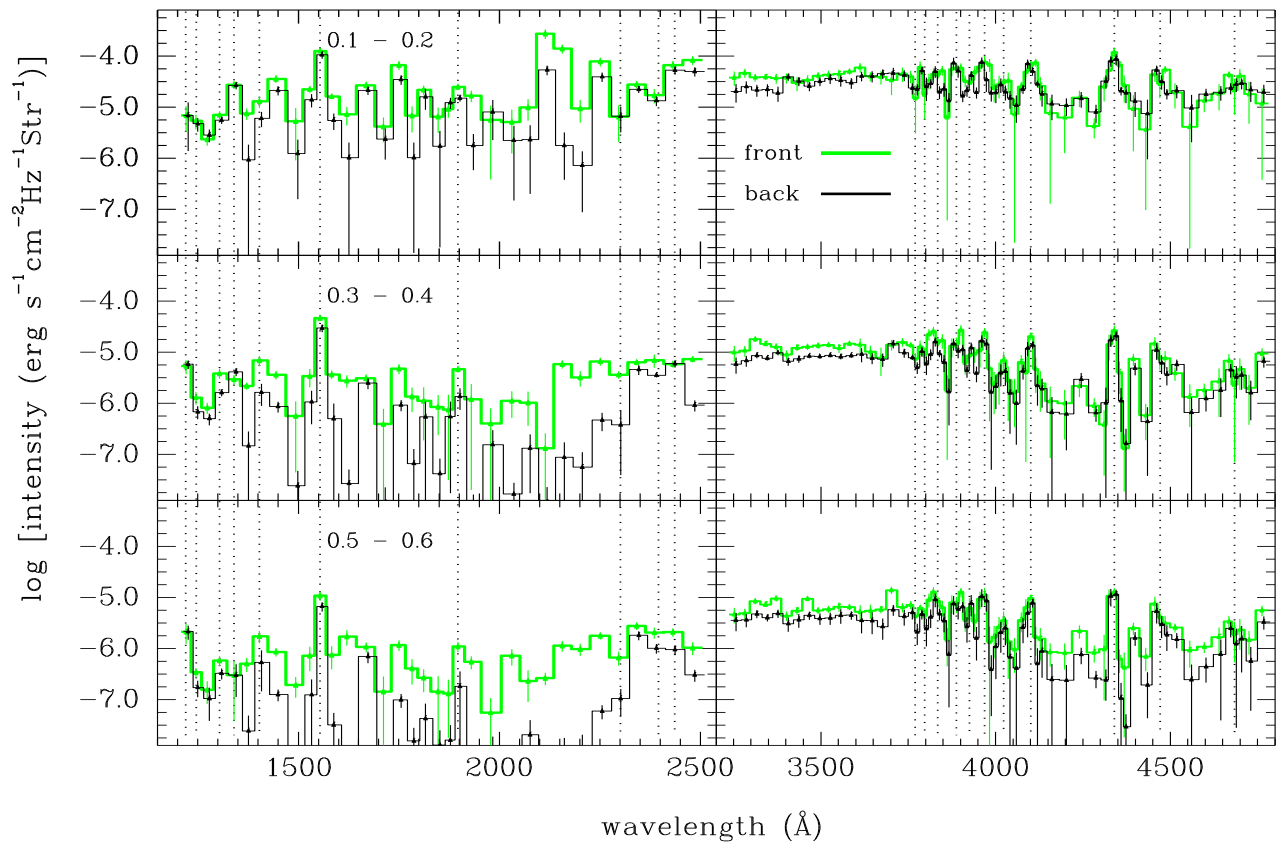}
     \caption{UV (left) and optical (right) spatially resolved disk
spectra for the azimuthal sections defined as front $(270^{o}-90^{o})$
(gray) and back $(90^{o}-270^{o};$ black lines) for three selected
annular regions (labeled in units of $R_{L1}$). Vertical dotted lines
mark the major transitions of the spectra.}
        \label{spec_disc}
  \end{figure*}

Our set of monochromatic eclipse maps allows us to separate the
spectra emitted by different parts of the accretion disk. Motivated by
the distinct emission observed in the gas stream region and by the
systematic difference in the emission from the opposite disk
hemispheres, we sliced the disk into three distinct regions:
``front'', ``back'' and ``gas stream''. Azimuths are measured from the
line joining both stars and increase counter-clockwise. We define
``front'' as the disk section between $270^{o}$ and $90^{o}$, and
``back'' as the region between azimuths $90^{o}$ and $270^{o}$. The
region defined as ``gas stream'' is depicted in Fig.\,\ref{mapa_esq}
(gray). In order to separate the disk spectra at different distances
from the disk center, we further divided each of the disk regions in
concentric annular sections of width $\Delta{R}=0.1R_{L1}$ .

In order to minimize the possible contributions from the bright spot
and gas stream to the disk ``front'' and ``back'' spectra we
calculated the symmetric disk-emission component in each
hemisphere. The symmetric component is obtained by slicing the disk
into a set of radial bins and by fitting a smooth spline function to
the resulting set of medians of the lower quartile of the intensities
in each bin. The spline fitted intensity in each annular section is
taken as the symmetric component. This procedure essentially preserves
the baseline of the radial profile, removing all azimuthal
structure. The statistical uncertainties affecting the fitted
intensities are estimated with the Monte Carlo procedure described in
Section 2.2.

The spatially resolved disk (front and back) spectra are shown in
Fig.\,\ref{spec_disc}. The inner disk shows a flat continuum which
becomes progressively fainter and redder with increasing radius,
indicating the existence of a radial temperature gradient. The lines
and the Balmer jump are in emission at all disk radii. The Balmer
decrement becomes flatter and the lines are more prominent and
narrower with increasing radii. The UV spectrum of the ``front'' side
is perceptibly different from that of the ``back'' side. The
comparison of the UV ``front'' and ``back'' spectra shows that the
latter contains broad absorption bands, possibly due to Fe\,II, that
become more conspicuous in the outer parts of the disk -- suggesting
that it arises from absorption along the line of sight by a vertically
extended region (e.g., Horne et al. 1994).

  \begin{figure*}
  \centering
  \includegraphics[bb=4cm 12cm 19cm 14cm,angle=-90,scale=.74]{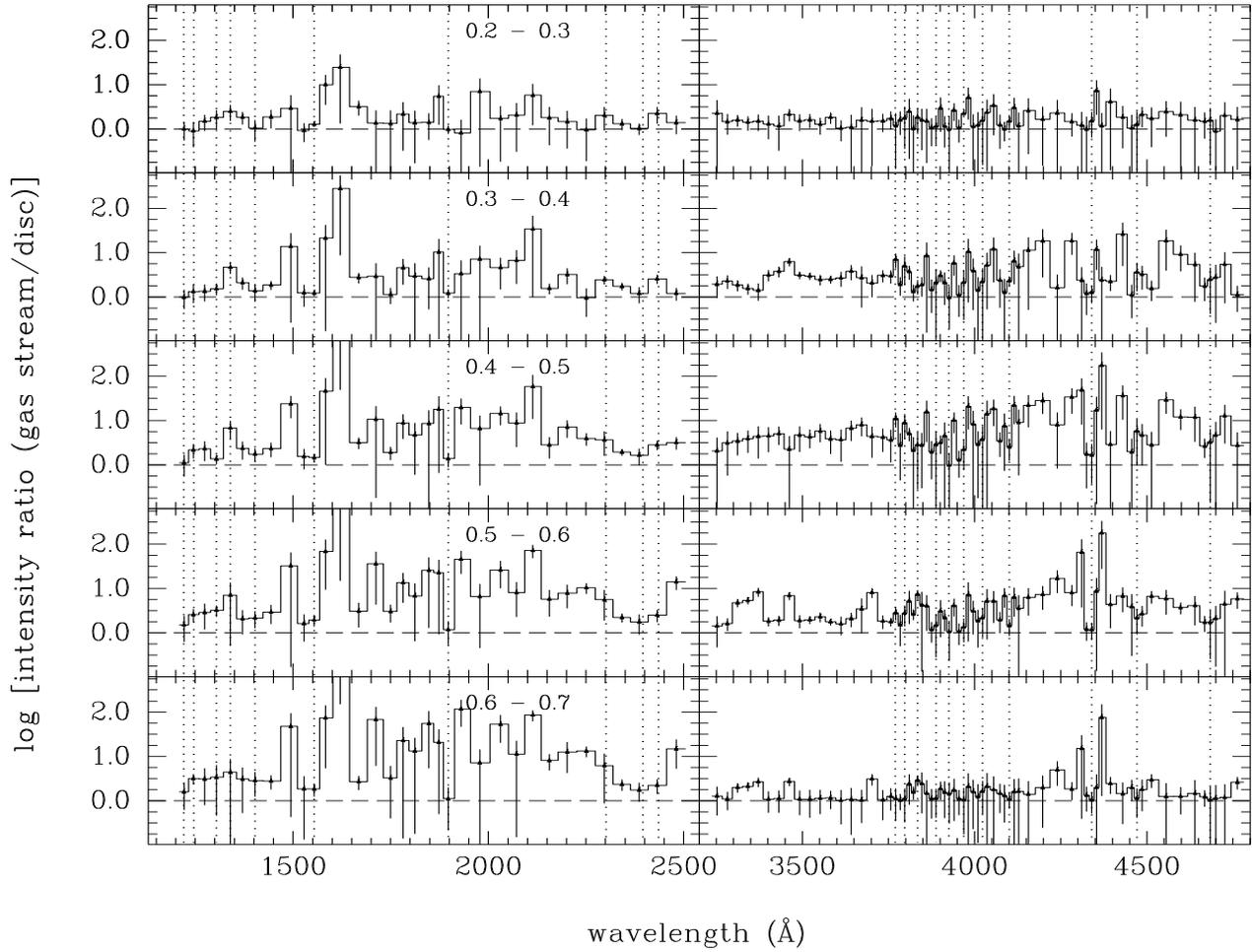}
     \caption{Ratio of the gas stream to the front disk spectra (see
Fig.\,\ref{mapa_esq}) as a function of radius. The notation is similar
to that of Fig.\,\ref{spec_disc}.}
        \label{spec_spot}
  \end{figure*}

In order to investigate the emission along the gas stream trajectory,
we calculate the ratio of the spectrum of the gas stream to the front
disk symmetric component at the same radius
(Fig.\,\ref{spec_spot}). The result reveals that the spectrum of the
gas stream is typically 1 order of magnitude brighter than the disk
spectrum in the intermediate and mainly in the outer disk regions
($0.3 \lesssim R/R_{L1} \lesssim 0.7$) and this difference is
significant at the $3\,\sigma$ level. This result suggests the
occurrence of gas stream overflow. This is in agreement with the
results of Baptista \& Bortoletto (2004), who found clear evidence of
gas stream overflow in B band eclipse maps of V2051 Oph in quiescence.

In contrast to what is observed in the disk spectra, in which the
lines appear in emission at all radii, in the ratio of the gas stream
to the disk spectrum the C\,IV, Si\,III $\lambda 1892$ and Balmer
lines appear in absorption. This suggests the presence of matter
outside of the orbital plane (e.g., chromosphere + wind) or that the
disk has a non-negligible opening angle (flared disk). Horne et
al. (1994) observed strong C\,IV emission out of the orbital plane in
the dwarf nova OY Car in quiescence and suggested that this emission
may arise from magnetic activity on the surface of the quiescent disk
(Horne \& Saar 1991) or from resonant scattering in a
vertically-extended region well above the disk.

The slope of the continuum, the strength of the Balmer jump and the
line intensities give a wealth of information about the physical
conditions in the emitting gas. It is clear that simple blackbody and
even stellar atmosphere models are not adequate to describe the strong
line emission spectra of the V2051 Oph accretion disk. A more
quantitative analysis of the spatially-resolved spectra demands
optically thin disk models. This is beyond the scope of the present
work and will be the subject of a forthcoming paper (A. Zabot et
al. 2006, in preparation).

\subsection{The emission lines}

\begin{figure*}
  \centering
  \includegraphics[bb=6cm 10cm 19cm 14cm,angle=-90,scale=.7]{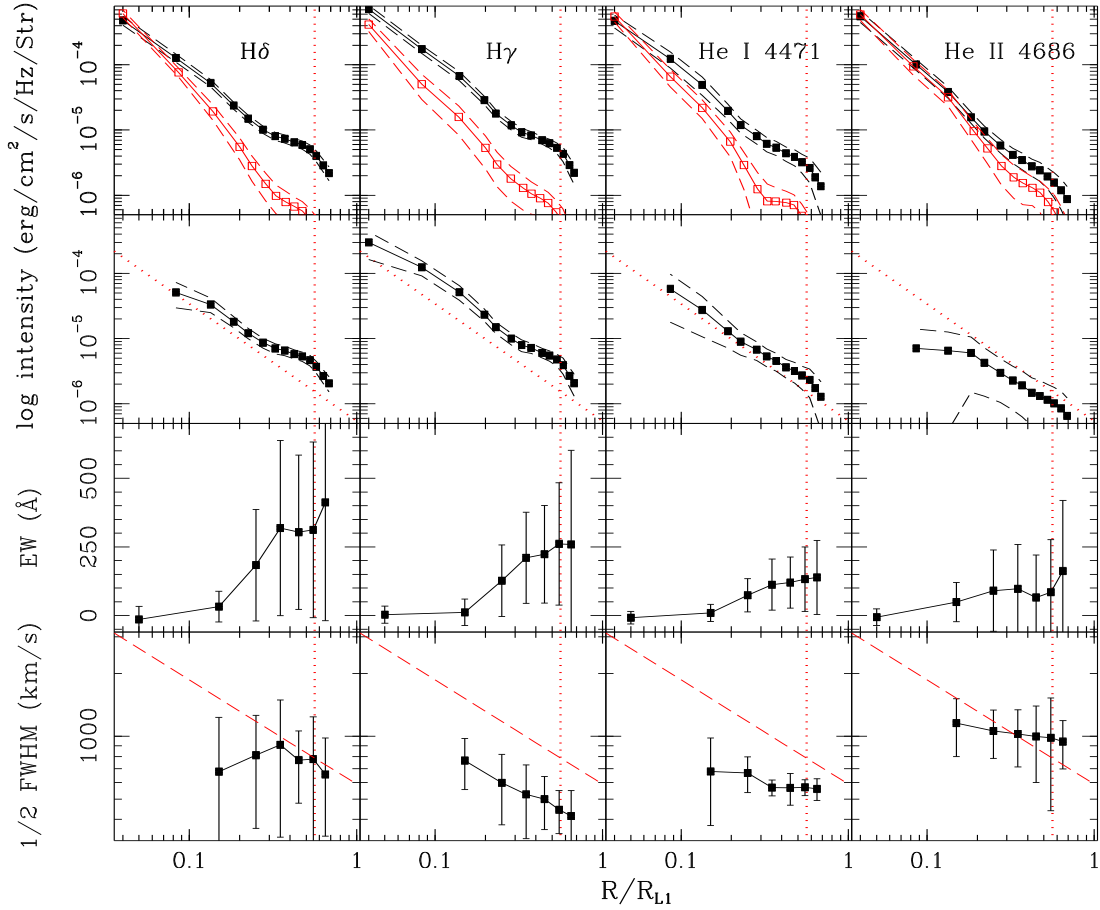}
     \caption{Top: Radial intensity distributions of selected optical
lines (filled symbols) and of the corresponding adjacent continuum
(open symbols). Dashed lines indicate the 1$\,\sigma$ limit in each
case. Dotted lines mark the outer disk radius of $R=0.56R_{L1}$
(Baptista et al. 1998). Second from the top: Net line emission radial
distributions. Dotted lines show the radial dependency for a slope $I
\propto R^{-1.78}$. Third from top: EW as a function of
radius. Bottom: FWHM of the lines as a function of radius. Dashed
lines show the law $v \propto R^{-1/2}$ expected for a gas rotating in
Keplerian orbits.}
        \label{ew_rad}
  \end{figure*}

In this section we analyze the radial behavior of selected emission
lines. Fig.\,\ref{ew_rad} shows the radial intensity distributions for
the lines and the adjacent continuum (top), the radial distributions
of the net line emission (second from top), the radial run of the
equivalent width (EW; third from top), and the radial run of the full
width at half-maximum (FWHM; bottom) for the H$\delta$, H$\gamma$,
He\,I 4471 and He\,II 4686 lines. The diagrams were computed for the
disk ``back'' region to avoid contamination by gas stream emission
and, as in Section 3.2, we used the symmetric disk-emission component
in this analysis. We remind the reader that our line eclipse maps
comprise the line emission plus any subjacent continuum
contribution. The line distributions were obtained from the average of
all eclipse maps along the line, while the continuum distributions
were obtained from the average of the nearest continuum maps on both
sides of each line. The net line emission distributions were computed
by the subtraction of the adjacent continuum from the corresponding
line distributions.

As expected, the intensity in the line maps is stronger than in the
adjacent continuum (indicating that the lines appear in emission),
except in the innermost disk regions, where the intensity of the
continuum reaches that of the lines.  For H$\delta$, He\,I 4471 and
He\,II 4686, the intensity in the line is equal or smaller than that
of the adjacent continuum in the innermost annulus (indicating that
the lines are in absorption in the white dwarf at the disk center) and
were not plotted in the corresponding net line emission panels. The
slope of the continuum distributions is steeper than that of the
lines. The net line distributions decrease in strength with increasing
disk radius. The lines are in emission at all disk radii with a radial
dependency $I \propto R^{-1.78\pm 0.06}$. This is steeper than the
empirical $I \propto R^{-1.5}$ law derived by Marsh et al. (1990)
assuming a Keplerian distribution of velocities for the line emitting
gas in the dwarf nova U Geminorum.

As a consequence of the radial behavior of the line and continuum
intensity distributions, the lines are relatively weaker in the inner
disk regions and their EW increases with increasing disk
radius. H$\delta$ and H$\gamma$ are the dominant lines with an EW
$\simeq 300$\,\AA\ at the disk edge ($\simeq 0.5\,R_{L1}$), while
He\,I 4471 and He\,II 4686 have EW $\simeq 100$\,\AA\ at the same
radius. In the innermost disk region ($R <0.1\,R_{L1}$) the EW becomes
negligible (or negative) because the continuum intensity reaches (or
exceeds) the line intensities. Because of this, the FWHM in this
region is also negligible and is not plotted in the corresponding FWHM
panels.

H$\delta$ and He\,II 4686 show FWHM values comparable to those
expected for a gas rotating in Keplerian orbits around an M$_{1}=0.78$
M$_{\sun}$ white dwarf, but the slope of the radial distribution is
flatter than the $v \propto R^{-1/2}$ law. Because of the large
uncertainties in the FWHM values, the distributions are still
consistent with the Keplerian expectation. On the other hand, the
H$\gamma$ and He\,I 4471 distributions are clearly different from the
Keplerian expectation. Both lines show sub-Keplerian velocities (at
the $2-3\,\sigma$ confidence level). While the slope of the H$\gamma$
distribution is consistent with the $v \propto R^{-1/2}$ law, He\,I
4471 shows a flat distribution, with velocities of 1/2 FWHM $\simeq
600-700\,km\,s^{-1}$ at all radii.

\subsection{The uneclipsed component}

The uneclipsed component was introduced in the eclipse mapping method
to account for the fraction of the total light that is not coming from
the accretion disk plane (e.g. light from the secondary star or from a
vertically-extended disk wind; Rutten et al. 1992).

\begin{figure*}
  \centering
  \includegraphics[bb=3cm 12cm 18cm 14cm,angle=-90,scale=.6]{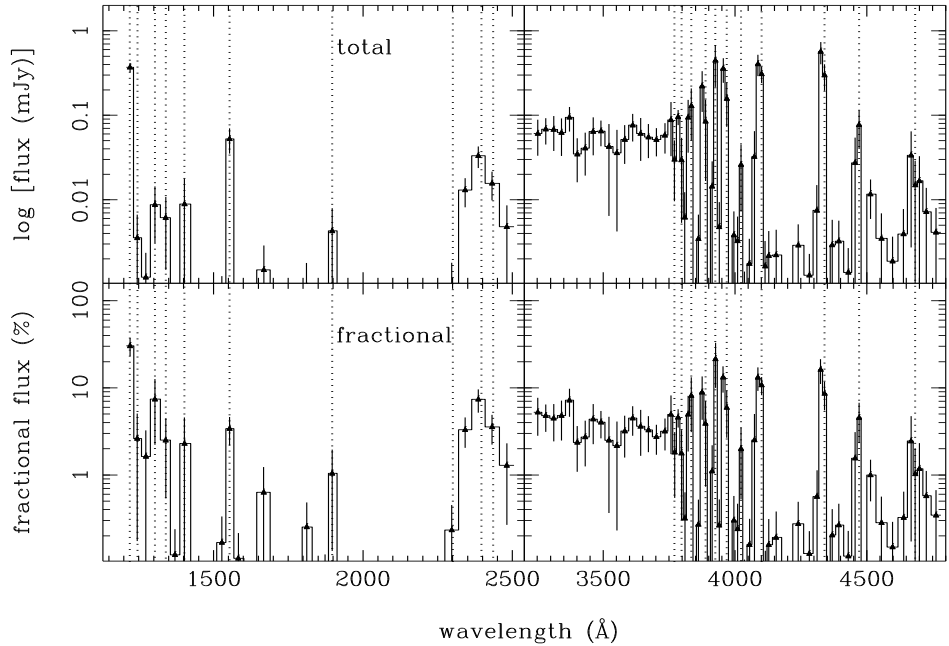}
     \caption{Spectrum of the uneclipsed component. Top: total
     contribution. Bottom: fractional contribution. The notation is
     similar to that of Fig.\,\ref{spec_disc}.}
        \label{neclip}
  \end{figure*}

Fig.\,\ref{neclip} shows the spectrum of the uneclipsed component, as
well as its fractional contribution, as a function of wavelength. We
estimated the fractional contribution of the uneclipsed component to
the total flux by computing the ratio of the uneclipsed light to the
average out of eclipse level at the corresponding passband.

The uneclipsed component is dominated by the Balmer continuum
emission, strong Balmer and He emission lines in the optical, and
resonant emission lines in the UV, suggesting that this light arises
in a hot, vertically extended, optically thin gas region.  $Ly\alpha$
is the strongest line in the uneclipsed spectrum ($\simeq 30$ per cent
of the total light at that wavelength). Nevertheless, most of this
contribution is of geocoronal origin and not intrinsic to V2051
Oph. The remaining strong lines in the optical and UV spectra, as well
as the Balmer continuum, account for $\simeq 10$ per cent of the total
light at the respective wavelength. The other continuum regions in the
uneclipsed spectrum yield negligible fractional contributions to the
total light. This indicates that the continuum emission comes mostly
from the orbital plane while part ($\simeq 10$ per cent) of the line
emission arises in a vertically extended region above the disk.

\section{Discussion}

\subsection{Evidences for chromospheric emission}

Our continuum eclipse maps, particularly in the UV
(Fig.\,\ref{mapas_uv}), show a front-back brightness asymmetry. This
becomes clear when we compare the spectrum of the front and back disk
sides (Fig.\,\ref{spec_disc}): the front side of the disk is brighter
than the back side. A similar result was found by Vrielmann et
al. (2002). They discussed a few possibilities (non-negligible disk
opening angle, warped disk, and bulge caused by impact of gas stream)
but found no compelling explanation for the effect.

Our front and back disk spectra were computed from a symmetric disk
component (see Section 3.2), which traces the baseline of the radial
intensity distribution and removes the contribution of asymmetric
brightness sources. Furthermore, the spectrum of the gas stream region
is markedly different from the disk spectrum at intermediate and large
radii ($R \gtrsim 0.3\,R_{L1}$). We are, therefore, confident that the
observed front-back asymmetry is not an artifact caused by azimuthal
smearing of bright spot emission, but it is a real effect.

Vrielmann et al. (2002) discussed the possibility that the observed
asymmetry could be caused by a non-negligible opening angle ($\alpha$)
in the accretion disk, but they discarded this hypothesis because it
would lead to an enhancement of the emission from the back side (seen
at the lower effective inclination $i_{eff}=i - \alpha$) in comparison
to the front disk side (seen at a higher effective inclination
$i_{eff}=i + \alpha$). This argument is only correct if the disk
atmosphere shows a stellar-type temperature decreasing with vertical
height. If the disk has a chromosphere, in which the temperature
actually increases with vertical height, the effect will be the
opposite. Because the front disk side is seen at higher angles, the
emerging photons will sample the uppermost (and hotter) chromospheric
layers and the resulting spectrum will be brighter. The back disk side
is seen through a lower inclination, allowing the emerging photons to
come from deeper (and cooler) layers and leading to fainter
intensities. Preliminary fits of the spatially resolved spectrum with
LTE atmosphere models lead to effective temperatures for the front
disk spectrum that are systematically higher than the temperatures of
the back disk spectrum at the same radius (A. Zabot et al. 2006, in
preparation), confirming the above scenario.

Therefore, the front-back asymmetry observed in the accretion disk of
V2051 Oph can be interpreted as evidence of chromospheric emission
from a disk with a non-negligible (but probably small) opening
angle. This chromosphere is the site of the emission lines and is
probably also responsible for the veiling of the continuum emission
from the compact underlying white dwarf and bright spot. It may also
account for the Fe\,II absorption features that become more pronounced
with increasing radius in the disk back side, because the photons
arising from these regions travel an increasingly larger path across
the vertically-extended disk chromosphere before leaving the binary
and reaching the observer.

\subsection{The white dwarf spectrum and the distance}

\begin{figure*}
  \centering
  \includegraphics[bb=3cm 3cm 19cm 23cm,,angle=-90,scale=.68]{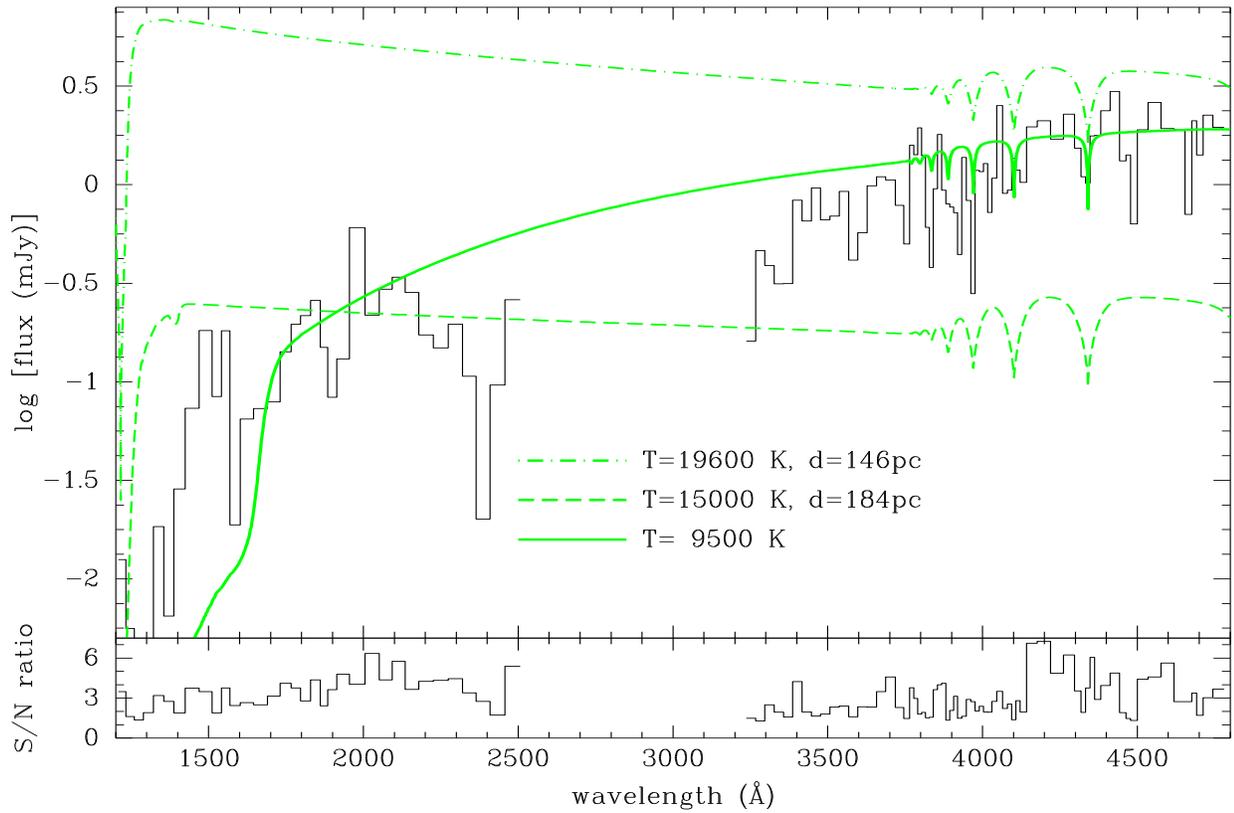}
     \caption{Spectrum of the V2051 Oph white dwarf (histogram) and DA
     white dwarf model fits. The dot-dashed line corresponds to a
     model with $T=19600 K, d=146\,pc$ (Vrielmann et al. 2002), and
     the dashed line correspond to a model with $T=15000 K, d=184\,pc$
     (Catal\'an et al. 1998). The best fit model is displayed as a
     solid line. The bottom panel shows the S/N of the the white dwarf
     spectrum.}
        \label{primary}
  \end{figure*}

Here we fit white dwarf atmosphere models to the white dwarf spectrum
in order to estimate the temperature of the primary star and the
distance to the object. We adopted a white dwarf radius of
$R_{WD}=0.0103\,R_{\odot}$ and a Roche lobe size of
$R_{L1}=0.422\,R_{\odot}$ (Baptista et al. 1998). Therefore, our
eclipse maps (see Section 2.3) have a scale factor of
$0.0392\,R_{L1}\,pixel^{-1}$. Since the primary comprises a diameter
of $0.0488\,R_{L1}$, the central pixel of the eclipse map is fully
contained in the white dwarf surface. We thus obtain a good estimate
of the white dwarf spectrum by extracting the flux of the central
pixel of the eclipse maps at each wavelength. These fluxes are
multiplied by a factor

\begin{equation}
\rm \frac{A_{WD}}{A_{pix}}=\left(
\frac{0.0244R_{L1}}{0.0392R_{L1}}\right)^{2} \frac {\pi}{cos \theta}
\hspace{2pt},
\end{equation}
to scale the spectrum to the effective area of the white dwarf. The
resulting spectrum shows a continuum filled with broad and shallow
absorption lines plus a Balmer jump in absorption, resembling that of
a DA white dwarf (Fig.\,\ref{primary}).

We employed a grid of DA white dwarf spectra with $6500\,K < T_{WD} <
20000\,K$ and $log~ g=8$ (D. Koester 2000, private communication) in
order to fit the V2051 Oph white dwarf spectrum. Since the DA white
dwarf models only account for the continuum level and the H\,I lines,
we masked the other line regions for the fitting procedure. This
includes removing the spectral region above 2300\,\AA~in the UV
spectrum and below 3400\,\AA~in the optical spectrum, to avoid
contamination by absorption bands due to Fe\,II. We considered two
possibilities: (i) the inner disk is opaque and the visible part of
the white dwarf surface is the projected area of the upper half
hemisphere above the opaque disk, (ii) the white dwarf surface is
fully visible through an optically thin disk.

Our best fit leads to a temperature of $T_{WD} =
9500\,^{+2900}_{-1900}\,K$ for the white dwarf with a distance of $d =
67\,^{+22}_{-25}\,pc$ if the inner disk region is opaque and $d =
92\,^{+30}_{-35}\,pc$ if the inner disk is optically thin. Both the
fitted temperature and the resulting distance are significantly
different from previously reported values.

Catal\'an et al. (1998) made a preliminary fit to the G160L extracted
white dwarf spectrum (assuming a white dwarf atmosphere model plus an
intervening cool gas layer to account for the Fe\,II absorption bands)
to find a white dwarf temperature of $15000\,K$. Their solution
provides a reasonable fit in the UV but underestimates the
white dwarf flux in the optical by a factor of 5. The white dwarf must
be cooler (and, therefore, the distance must be smaller) to match the
slope of the combined UV-optical spectrum. On the other hand,
Vrielmann et al. (2002) fitted white dwarf atmosphere models to UBVRI
photometric measurements to find $T_{WD} = 19600\,K$ (if only the
upper half of the white dwarf is visible) or $T_{WD} = 15000\,K$ (if
the white dwarf is fully visible) and $d = 146\,pc$. However, their
result largely overestimates the white dwarf contribution in the
UV. For comparison, Fig.\,\ref{primary} shows DA white dwarf
models for the temperatures and respective distances obtained by the
works mentioned above.

As a first step in trying to understand the difference, we investigate
the possibility that some systematic error in the eclipse-mapping
analysis (e.g. wavelength dependent blurring of bright central
sources) is affecting the slope of the extracted white dwarf
spectrum. To test this hypothesis, we extract the white dwarf spectrum
by measuring the height of the jump in flux at white dwarf egress in
the set of narrow band UV and optical light curves used for eclipse
mapping. This provides an estimate of the white dwarf flux as seen
during egress and is quite similar to the procedure used by Catal\'an
et al. (1998) to extract the UV white dwarf spectrum. The resulting
spectrum is consistent with that of Fig.\,\ref{primary} under the
uncertainties. We are therefore confident that the red slope of the
white dwarf spectrum of V2051 Oph at the time of the HST observations
is real. Thus, the main point is that, it is not possible to model the
observed white dwarf spectrum with a hot ($T_{WD} \gtrsim 14000\,K$)
DA atmosphere model.

The origin of the discrepancy between our result and the previous
analyses may be in the restricted spectral coverage of the data used
in those works. Catal\'an et al. (1998) used only the UV part of the
spectrum ($1150 - 2500\,\AA$) to fit the white dwarf temperature. It
is possible to match the observed UV intensities with a DA spectrum of
$15000\,K$ for a distance of $d = 184\,pc$ (Vrielmann et al. 2002) at
the expense of a large mismatch in the optical.

The slope of a DA white dwarf spectrum is very similar in the optical
($3500 - 5000\,\AA$) for a large range of temperatures. Therefore, the
optical spectrum alone does not provide a strong constraint to the
white dwarf temperature. We tested this by fitting white dwarf models
only to the optical part of the spectrum. This exercise shows that it
is possible to fit temperatures in the range $6500 - 16500\,K$ with a
$\chi ^{2}$ only $20\%$ larger than the $\chi ^{2}$ of the best-fit
model in this case, $T_{WD} = 11500\,K$. This lack of sensitivity
affects the white dwarf temperature determination of Vrielmann et
al. (2002), since they used only optical photometric data in their
analysis. Moreover, they mistakenly took the value
$R_{WD}=0.0244\,R_{\odot}$ as the white dwarf radius, while the value
quoted by Baptista et al. (1998) is $R_{WD}=0.0103\,R_{\odot}$. To
compensate for the larger radius assumed for the white dwarf, their
fit had to increase the distance by a similar amount (a factor of
$\sim 2.3$).

Steeghs et al. (2001) used optical spectroscopy $(4000 - 6800\,\AA)$
to find that a $(15000\,K)$ blackbody was a good match to the slope of
their extracted white dwarf spectrum. Their result faces the same
problem discussed above; the slope of the optical continuum is very
similar for a large range of temperatures.

It is important to notice that at the time of our observations V2051
Oph was in an unusual low brightness state, in which mass accretion
may have been considerably reduced (Baptista et al. 1998). With
reduced (or absent) accretional heating, it is possible for the white
dwarf to cool down. We therefore expect that our fitted white dwarf
temperature is lower than that during a normal, quiescent state.

Our larger spectral coverage, combining optical and UV data, allows a
better determination of the slope of the white dwarf spectrum, and,
therefore, its temperature and distance. Our results show that the
distance to V2051 Oph is smaller than previously found.


\section{Conclusions}

The main results of our spectroscopic study of V2051 Oph in a faint
brightness state during 1996 can be summarized as follows:

\begin{enumerate}

\item The presence of white dwarf and bright-spot strong emission in
the continuum maps and their absence in the line maps, coupled with
the significant extra absorption in the spectra of the back disk side,
are indications of the vertical extension and large optical depth of
the gas from which the lines originate.

\item Distinct emission along the stream trajectory suggests the
occurrence of gas stream overflow.

\item Spatially resolved spectra show that the lines are in emission
 at all disk radii. The Balmer decrement becomes shallower with
 increasing radius.

\item The FWHM of the emission lines differs from that expected for a
gas in Keplerian rotation and the line intensities drop with a radial
dependency of $I \propto R^{-1.78\pm 0.06}$.

\item The spectrum of the uneclipsed light is dominated by strong
 emission lines and a Balmer jump in emission, indicating origin in a
 hot, vertically extended, optically thin gas region above the
 disk. The strongest uneclipsed lines contribute $\simeq 10$ per cent
 of the total flux.

\item Strong absorption bands, possibly due to Fe II, are seen in the
 spectra of the back disk side, suggesting it arises from absorption
 by an extended gas region above the disk.

\item The front disk spectrum is systematically brighter than the back
disk spectrum at the same radius. This can be explained in terms of
chromospheric emission (higher temperatures at the uppermost
atmospheric layers) from an accretion disk with a non-negligible
opening angle (limb brightening effect).

\item We fit stellar atmosphere models to the extracted white dwarf
 spectrum to find a temperature $T_{WD} =
 9500\,^{+2900}_{-1900}\,K$ and a distance of $d =
 67\,^{+22}_{-25}\,pc$ (if the inner disk is opaque) or $d =
 92\,^{+30}_{-35}\,pc$ (if the inner disk is optically thin).

\end{enumerate}

\acknowledgments

{\it Acknowledgments.} We thank the anonymous referee for useful
comments and suggestions. The white dwarf atmosphere models used in
this work were kindly provided by Detlev Koester. This work was
partially supported by CNPq/Brazil through the research grant
62.0053/01-1 -- PADCT III/Milenio. RB acknowledges financial support
from CNPq/Brazil through grants n. 300.354/96-7 and
301.442/2004-5. RKS acknowledges financial support from CAPES/Brazil
and CNPq/Brazil.

\appendix

\section{Reconstruction of maps from light curves with incomplete phase coverage}

Here we address the reliability of eclipse-mapping brightness
reconstructions for the case of light curves with incomplete phase
coverage, such as those of run H1.

\begin{figure}
   \centering \includegraphics[bb=1cm 2cm 20cm 18cm,scale=.6]{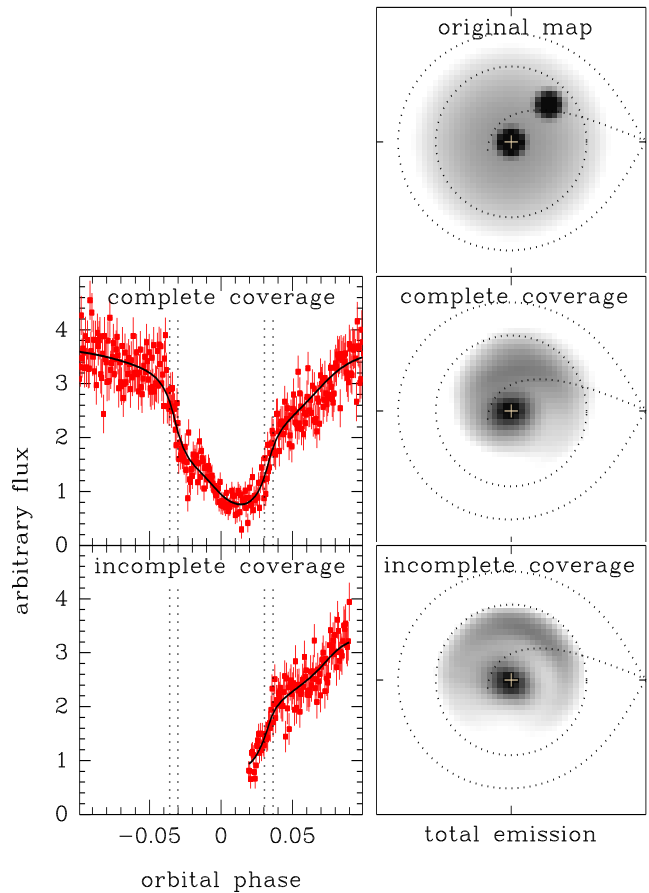}
      \caption{Left: Synthetic light curves with S/N=10 for the full
      phase coverage (top) and for the phase set of the optical run
      (bottom). Right: Original map and reconstructions obtained from
      the light curves in the left panels. The notation is the same as
      in Fig.\,\ref{mapas_uv}.}
         \label{simul}
   \end{figure}

In order to simulate the structures observed in the optical maps, we
created artificial eclipse maps of $51 \times 51$ pixels with two
compact Gaussian spots, one at the disk center (the white dwarf) and
another near the impact region of the gas stream with the disk (the
bright spot), on top of a low-intensity, broad Gaussian background to
simulate a faint accretion disk. The artificial map is shown in
Fig.\,\ref{simul}.

We simulated the eclipse of the artificial map using the geometry of
V2051 Oph ($i=83\degr$ and $q=0.19$, Baptista et al. 1998) to create
light curves with added Gaussian noise reproducing the S/N of the
V2051 Oph data ($S/N=10$). Since the optical light curves of V2051 Oph
have incomplete phase coverage, we truncated the synthetic light
curves to reproduce the set of orbital phases of the optical
run. Fig.\,\ref{simul} shows light curves for a complete phase
coverage and for the set of phases of the V2051 Oph optical data. We
applied eclipse mapping techniques to the two light curves and
compared the results with the original map.

\begin{figure}
   \centering \includegraphics[bb=1cm 9cm 20cm 23cm,scale=.5]{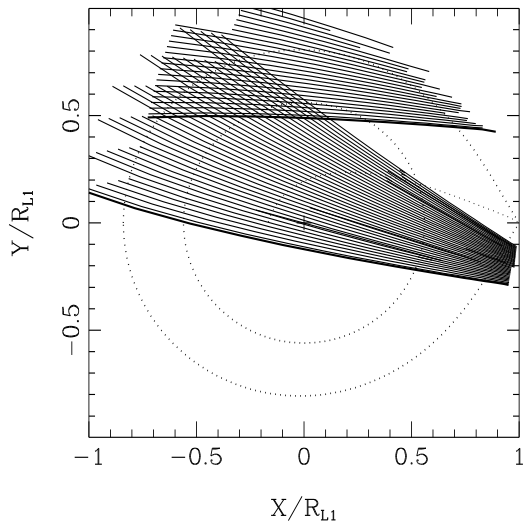}
      \caption{Eclipse geometry of V2051 Oph for the set of phases of
the optical run.  Arches (solid lines) connect points of constant
ingress and egress phases. To avoid overcrowding the figure, only one
phase/arch of every three is shown. The pair of thick lines shows the
ingress and egress arches for the first phase of the light curve
($\phi= +0.0194$).  Dotted lines depict the primary Roche lobe, a disk
of radius $0.56\;R_{L1}$ and the ballistic stream trajectory.}
         \label{arcos}
   \end{figure}

The ability of the eclipse-mapping method to construct a
two-dimensional surface brightness map from a one-dimensional eclipse
light curve stems from the fact that ingress and egress arches
intersect with each other at relatively large angles (Horne
1983). Fig.\,\ref{arcos} illustrates the eclipse geometry of V2051 Oph
for the incomplete set of phases of run H1. Arches drawn on the
eclipse map connect points of constant ingress and egress phases. The
pair of thick lines shows the ingress/egress arches for the first
phase of the light curve ($\phi= +0.0194$).  Each pair of arches
outlines a region of the disk that is being occulted by the shadow of
the secondary star at a given phase (for visualization purposes,
Fig.\,\ref{arcos} shows only one of every three phases/arches in the
incomplete light curve).

The effect of incomplete phase coverage is to reduce the information
available to reproduce structures in parts of the accretion disk.  For
the light curve with incomplete phases, about 40 \% of the disk
surface is not covered by ingress or egress arches
(Fig.\,\ref{arcos}). There is no information in the eclipse shape
about the brightness distribution of the disk regions not sampled by
the grid of arches. Therefore, the disk side approaching the secondary
star (the lower hemisphere of the map in Fig.\,\ref{simul}) will not
affect the shape of the eclipse light curve with the incomplete set of
phases and its brightness distribution will not be recovered by the
eclipse-mapping method.

Fig.\,\ref{simul} shows that the structures are reasonably well
reconstructed with the light curve of full eclipse phase coverage. As
a consequence of the intrinsic azimuthal smearing effect of the
eclipse mapping technique (see Baptista 2001 for more details) the
asymmetric spot appears smeared into a crescent-shaped
structure. Because the light curve with incomplete phase coverage
lacks most of the ingress phases, the technique is not able to
properly reconstruct structures located in the disk region that
approaches the secondary. However, for this V2051 Oph simulation the
main brightness structures are in the disk center and in the bright
spot region (the disk hemisphere moving away from the secondary star),
and the reconstruction is reasonably good and quite similar to that
obtained with the complete eclipse phase coverage. The radial
intensity distributions of the eclipse maps from the complete and
incomplete phase coverage light curves (not shown) are very similar,
with only small differences that are within the uncertainties of the
distributions.
 
In summary, the simulations show that, although the eclipse mapping
method is not able to recover structures in disk regions for which
there is no information in the eclipse shape, brightness distributions
with structures in the disk center and in the bright spot region can
still be detected in eclipse maps derived from light curves with an
incomplete phase coverage such as that of run H1, in spite of their
relatively low S/N.

\end{document}